\documentclass[11pt,a4paper]{article}
\pdfoutput=1
\usepackage{imports}
\usepackage{generalmacros}
\usepackage{pgfplots}

\newcommand{\C}{\mathcal{C}}
\newcommand{\happiest}{\ensuremath{h^+}}

\newcommand{\instance}{\ensuremath{(N, \mathcal C, \mathcal V)}}
\newcommand{\myinstance}{\textup{\textrm{SANN}}}
\newcommand{\diameter}{\operatorname{diameter}} \pgfplotsset{compat=1.18}

\begin{document}
\title{Connected Equitable Cake Division via Sperner's Lemma}
\date{}

\author[1]{Umang Bhaskar}
\author[2]{A. R. Sricharan\textsuperscript{\dag}}
\author[3]{Rohit Vaish}
\affil[1]{Tata Institute of Fundamental Research, India; \texttt{\small{umang@tifr.res.in}}}
\affil[2]{Faculty of Computer Science, Doctoral School of Computer Science,
University of Vienna, Austria; \texttt{\small{sricharana67@univie.ac.at}}}
\affil[3]{Indian Institute of Technology Delhi, India; \texttt{\small{rvaish@iitd.ac.in}}}
\date{}

\renewcommand*{\thefootnote}{\fnsymbol{footnote}}
\footnotetext[2]{This work was done when the second author was at Chennai Mathematical Institute.}
\renewcommand*{\thefootnote}{\arabic{footnote}}

\maketitle

\begin{abstract}
We study the problem of fair cake-cutting where each agent receives a connected piece of the cake. A division of the cake is deemed fair if it is \emph{equitable}, which means that all agents derive the same value from their assigned piece. Prior work has established the existence of a connected equitable division for agents with nonnegative valuations using various techniques. We provide a simple proof of this result using Sperner's lemma. Our proof extends known existence results for connected equitable divisions to significantly more general classes of valuations, including nonnegative valuations with externalities, as well as several interesting subclasses of general (possibly negative) valuations.
\end{abstract}

\section{Introduction}
\label{sec:intro}

Cake-cutting is a fundamental problem at the intersection of economics, political science, and computer science~\citep{BT96fair,RW98cake,M04fair,BCE+16handbook}. The problem involves a divisible, heterogeneous resource, often called a ``cake'', that should be divided fairly among agents with differing preferences. Over the years, the cake-cutting problem has generated significant theoretical interest, leading to intriguing connections with various areas of mathematics~\citep{DS61cut,A87splitting,MBZ+03using}. Additionally, the problem has found practical applications in modeling fair allocation of desirable resources like land estates~\citep{S17fair} and time slots~\citep{HIS20fair} as well as undesirable resources such as rent~\citep{Su99sperner}.

In this work, we focus on a fairness notion called \emph{equitability}~\citep{DS61cut}, which requires that each agent derives the same level of utility from the portion of the resource assigned to it. This notion captures fairness from the perspective of a central planner aiming to minimize the disparity between the best-off and worst-off agents. Numerous studies have explored the existential and computational aspects of equitability and its connections with other fairness concepts~\cite{DS61cut,A87splitting,BJK06better,CP12computability,CDP13eqdivisible,AD15connected,PW17lower}. Experiments with human participants have shown that equitability is a more accurate predictor of perceived fairness in resource allocation compared to other concepts such as envy-freeness~\citep{HP09envy}.\footnote{A division is \emph{envy-free} if each agent values the resource assigned to it at least as much as that assigned to any other agent~\citep{F66resource}.} Equitability is also a key feature of the well-known \emph{Adjusted Winner} procedure~\citep{BT96fair}.

Formally, a cake is represented by the interval $[0,1]$. A cake division refers to a partition of the interval $[0,1]$ among $n$ agents; each agent's assigned part is called its ``piece''. Given any cake division, each agent derives a value from its assigned piece, which is specified by its \emph{valuation function}.
A seminal result by Dubins and Spanier~\cite{DS61cut} showed that an equitable cake division always exists for any given \emph{additive} valuation functions. However, in this case, an agent's piece can be any member of the $\sigma$-algebra of subsets of $[0,1]$. Such divisions can be impractical in certain situations, particularly when dividing land or allocating time slots. As Stromquist~\citep{S80cut} memorably described it: ``A player who hopes only for a modest interval of cake may be presented instead with a countable union of crumbs.''

Motivated by these considerations, our work aims to study the existence of a \emph{connected} equitable cake division. In this type of division, each agent receives a disjoint subinterval of $[0,1]$, and all agents derive the same value from their respective pieces. Interestingly, there are several proofs demonstrating the existence of a connected equitable cake division~\citep{AD15connected,CDP13eqdivisible,Ch17eqdivisible, SHS18}. However, these proofs either apply to a restricted class of valuation functions or use sophisticated techniques and are, therefore, quite complex.

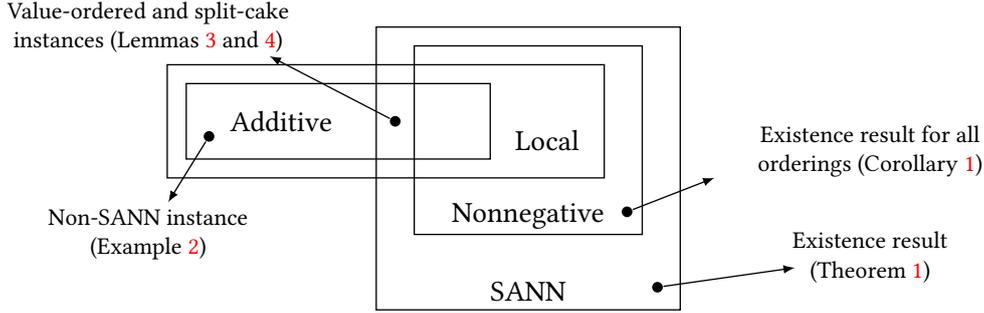
\begin{figure}[t]
    \centering
	\tikzset{every picture/.style={line width=0.5pt}}
	\begin{tikzpicture}
		\footnotesize
		\draw (0,0) rectangle (4,1);
		\node (1) at (1.25,0.5) {\normalsize{Additive}};
		\draw (-0.25,-0.25) rectangle (5.5,1.25);
		\node (1) at (4.75,0.25) {\normalsize{Local}};
		\draw (3,-1) rectangle (6,1.5);
		\node (1) at (4.5,-0.75) {\normalsize{Nonnegative}};
		\draw (2.5,-2) rectangle (6.5,1.75);
		\node (1) at (4.5,-1.75) {\normalsize{\myinstance{}}};
            \node[circle,fill=black,minimum size=4pt,inner sep=0pt] (1) at (0.3,0.3) {};%
            \draw (-0.5,-1) node[align=center] (2) {{Non-\myinstance{} instance}\\{(Example~\ref{fig:non-example})}};%
            \draw[shorten >=0.5cm,-latex] (1) -- (2.center);%
            \node[circle,fill=black,minimum size=4pt,inner sep=0pt] (1) at (5.8,-0.7) {};%
            \draw (9,0.1) node[align=center] (2) {{Existence result for all}\\{orderings (Corollary~\ref{cor:indexing})}};%
            \draw[shorten >=2.2cm,-latex] (1) -- (2.north);%
            \node[circle,fill=black,minimum size=4pt,inner sep=0pt] (1) at (6.2,-1.7) {};%
            \draw (9,-1.3) node[align=center] (2) {{Existence result}\\{(Theorem~\ref{thm:existence})}};%
            \draw[shorten >=1cm,-latex] (1) -- (2.center);%
            \node[circle,fill=black,minimum size=4pt,inner sep=0pt] (1) at (2.75,0.5) {};%
            \draw (-0.5,1.75) node[align=center] (2) {{Value-ordered and split-cake}\\{instances (Lemmas~\ref{lem:value-ordered} and~\ref{lem:split-cake})}};%
            \draw[shorten >=1.8cm,-latex] (1) -- (2.north);%
	\end{tikzpicture}
    \caption{Summary of our results in the form of a Venn diagram. Each rectangle denotes a class of instances~(defined in Section~\ref{sec:prelim}) and each dot-arrow pair denotes the domain on which the corresponding result applies.}
    \label{fig:Summary}
\end{figure}

\paragraph{Our Contributions.} We make the following technical contributions (also see Figure~\ref{fig:Summary}):
\begin{itemize}
    \item We define a new class of instances that we call \emph{some agent nonnegative} (\myinstance{}) instances (see Section~\ref{sec:cakecutting}). Essentially, in a \myinstance{} instance, any connected division results in some agent receiving a piece of cake with nonzero size for which it has nonnegative value. This property crucially depends on the ordering of the agents. However, this class allows valuations with externalities (we call these \emph{global} valuations) where the value derived by agent $i$ under a given connected division can depend on \emph{all} pieces, not just the piece given to agent $i$. Our results also extend directly to \emph{some agent nonpositive} (SANP) instances where any connected division results in some agent receiving a piece of cake with nonzero size for which it has nonpositive value.
    \item In Section~\ref{sec:existence}, we present a simple proof of the existence of a connected equitable division for \myinstance{} instances (Theorem~\ref{thm:existence}). Our proof uses Sperner's lemma~\cite{S28neuer}, which is a combinatorial counterpart of Brouwer's fixed point theorem and has been used in various applications such as in finding competitive equilibria in an exchange economy~\citep{S67core}, envy-free cake-cutting, and envy-free rent division~\citep{Su99sperner}. To the best of our knowledge, this is the first application of Sperner's lemma in the context of equitability.

    \item We identify several interesting subclasses of \myinstance{} instances that demonstrate how our work generalizes known results. Prior work has shown that under \emph{additive} nonnegative valuations, \emph{any} given ordering of agents admits a connected equitable division. We generalize this result to \emph{all} nonnegative valuations (Corollary~\ref{cor:indexing}). We also explore \emph{possibly negative} additive valuations. Here, we show that commonly-studied settings such as identical valuations~(and its generalization, value-ordered instances) as well as single-peaked valuations~(and its generalization, split-cake instances) are included in the \myinstance{} class~(see Section~\ref{sec:nonmonotone}), and thus permit a connected equitable cake division for a particular agent ordering.
\end{itemize}

\subsection*{Related Work}

The existence of connected equitable divisions of a cake has been studied in a number of previous papers. Aumann and Dombb~\cite{AD15connected} showed that there is an equitable division that maximizes the egalitarian welfare among all connected divisions using the compactness of the space of all connected divisions. Their proof relied on the valuations being nonnegative and continuous. By contrast, our proof of existence of a connected equitable division applies to a broader class of \emph{global} valuations, where an agent's value for a division can
depend not only on the position of the cut points of the agent's piece, but also on the position of the cut points of other pieces. Furthermore, our proof applies to subclasses of possibly negative valuations, a generalization not allowed by Aumann and Dombb's proof.

In a separate study, Cechl\'{a}rov\'{a} et al.~\cite{CDP13eqdivisible} presented an existence proof based on function inverses. This proof crucially required the valuations to be defined as integrals of nonnegative density functions. The proof also showed that if all density functions are strictly positive, then for any instance, there is a unique connected equitable division.

A simpler proof of existence by Ch{\`e}ze~\cite{Ch17eqdivisible} proceeds via another fixed-point result, the Borsuk-Ulam theorem~\citep{MBZ+03using}. While this result is stated for valuations defined in terms of nonnegative density functions, it extends to %
more general valuations that may depend on the pieces allocated to other agents as well.
The use of Borsuk-Ulam requires antipodal functions (i.e., functions $f$ for which $f(x) = - f(-x)$), and cannot directly handle valuations that may be negative. Specifically,  when the valuations are allowed to be negative, not every %
fixed point of the function used by Ch{\`e}ze corresponds to an equitable division. On the other hand, our proof via Sperner's lemma readily shows the existence for many classes of possibly negative valuations as well.

For agents with \emph{identical} valuation functions, Avvakumov and Karasev~\cite{AK23identicalef} showed the existence of a connected equitable division via an involved topological argument. Their proof applies to valuations defined on intervals that are continuous in the two interval end-points, and hence is quite general. While our result is less general in comparison (since we obtain the existence for identical additive valuations defined via density functions), our technique is significantly simpler.

Cechl{\'{a}}rov{\'{a}} and Pill{\'{a}}rov{\'{a}}~\cite{CP12computability} studied algorithms for computing connected equitable divisions that are also \emph{proportional}; that is, each agent's value for its piece is at least $1/n$ fraction of its value for the entire cake. They considered nonnegative valuations and show that even for three agents, there is no finite algorithm for obtaining a proportional and equitable connected division. However, considering approximate equitability %
allows for an efficient algorithm; specifically, their algorithm requires $\mathcal{O}\!\left( n \log (n/\eps) \right)$ queries in the Robertson-Webb model.

Procaccia and Wang~\cite{PW17lower} studied the query complexity of computing an (approximately) equitable cake division without the connectedness constraint in the Robertson-Webb query model~\cite{RW98cake}. They showed that no finite algorithm exists for finding an equitable division, even without the connectedness constraint. They also studied the query complexity of $\eps$-equitability, where the absolute difference between the highest and lowest utilities is at most $\eps$. For this problem, they established upper and lower bounds of $\mathcal{O} \left( (1/\eps) \ln (1/\eps) \right)$ and $\Omega \left( \ln (1/\eps) / \ln \ln (1/\eps) \right)$, respectively. For the case of two agents, Br{\^{a}}nzei and Nisan~\cite{BranzeiN22} provided a lower bound of $\Omega \left( \ln (1/\eps) \right)$, which matches the upper bound of $\mathcal{O} \left( \ln (1/\eps) \right)$ due to Cechl{\'{a}}rov{\'{a}} and Pill{\'{a}}rov{\'{a}}~\cite{CP12computability} for connected approximately equitable divisions.

Finally, we will touch on the literature related to envy-freeness, which requires that every agent prefers its own piece to that of any other agent. It is known that a ``perfect'' division, in which each piece has the same value for every agent, always exists when connectedness is not required~\cite{DS61cut,W80dividing,A87splitting}. Such a division is both envy-free and equitable. However, under the connectedness constraint, it has been shown that envy-freeness is not compatible with equitability, even with just three agents~\citep{BJK06better}. Notably, Su~\citep{Su99sperner} used Sperner's lemma to demonstrate the existence of a connected envy-free division under a large class of valuations that subsumes nonnegative global valuations (this proof is credited to Forest Simmons). Our proof for the existence of a connected equitable division is inspired by Su's construction. It is relevant to note that Su's construction works for triangulations that admit an ownership assignment, wherein each vertex of the triangulation is assigned an index such that no elementary triangle gets a repeated label. By contrast, our proof works with all triangulations of the simplex and not just ones that admit an ownership assignment. Connected and envy-free cake cutting under possibly negative valuations has also been studied in prior work~\cite{SH18burntcake,MZ19burntprime,AK21primepowers}.

\section{Preliminaries}
\label{sec:prelim}

We denote $\{ 1, \ldots, n \}$ as $[n]$. We use $e_i$ for the unit vector with $1$ in the $i$th coordinate and $0$ in all other coordinates.

\subsection{Sperner's Lemma}
\label{sec:sperners}

\paragraph{$n$-simplex.}
An $n$-\emph{simplex} is the convex hull of $n+1$ independent vectors. %
That is, for independent vectors $u_1, u_2, \ldots, u_{n+1}$, the corresponding $n$-simplex is given by the set of points
$$ \delta \coloneqq \left\{ \sum_{i=1}^{n+1} \alpha_i u_i \mid \alpha_i \ge 0 \text{ for all } i \text{ and } \sum_i \alpha_i = 1 \right\}.$$
The extreme points of the simplex, namely $u_1,u_2,\dots,u_{n+1}$, are called the \emph{vertices} of the simplex.
The \emph{standard} $n$-simplex $\Delta^n$ is given by the standard vectors $u_i = e_i \in \R^{n+1}$. That is,
$$\Delta^n = \left\{ x \in \R^{n+1} \mid \sum_i x_i = 1, x_i \ge 0 \right\}.$$

\paragraph{$k$-face.}
A $k$-\emph{face} of an $n$-simplex $\delta$ is the $k$-simplex formed by the convex hull of any subset of $k+1$ vertices. For any $S \subseteq \{1,2,\dots,n+1\}$, the face spanned by $S$ is given by $$F_S \coloneqq \left\{ \sum_{i=1}^{n+1} \alpha_i u_i \mid \alpha_i \ge 0 \text{ for all } i, \alpha_i = 0 \text{ for all } i \notin S, \text{ and } \sum_i \alpha_i = 1 \right\}.$$
For example, the side opposite vertex $u_i$ of the simplex, $F_{[n+1] \setminus \{i\}}$, is an $(n-1)$-face. Such $(n-1)$-faces for any vertex are called \emph{facets}.

\paragraph{Triangulation.}
A \emph{triangulation} $T$ of the standard $n$-simplex $\Delta^n$ is a finite collection of $n$-(sub)simplices whose union is $\Delta^n$, with the property that any two of them intersect in a face common to both, or not at all. Formally, a collection of $n$-simplices $\{\delta_i\}_{i \in \mathcal I}$ is a triangulation of $\Delta^n$ if $\cup_i \delta_i = \Delta^n$, and for all $i, j \in \mathcal I$, either $\delta_i \cap \delta_j = \emptyset$, or there exists a face $F$ of $\delta_i$ such that $\delta_i \cap \delta_j = F$.

The smaller simplices are called \emph{elementary simplices}, and the union of the vertices of all elementary simplices are called \emph{vertices of the triangulation}.

\paragraph{Labeling of a triangulation.}
Given a triangulation $T$ of a standard $n$-simplex $\Delta^n$, a \emph{labeling} of $T$ is a function that gives each vertex of the triangulation a label from $\{1, 2, \ldots n+1 \}$.
A labeling is called a \emph{Sperner labeling} if it satisfies the following two properties:
\begin{itemize}
    \item Each vertex of the simplex $\Delta^n$ has a different label.
    \item The facet opposite the vertex of $\Delta^n$ labeled $i$ has no vertex of the triangulation labeled $i$.
\end{itemize}

An elementary simplex is called \emph{fully labeled} if all of its vertices have different labels.

\begin{restatable}[Sperner's lemma;~\citealp{S28neuer}]{prop}{Sperners}
\label{lem:Sperners}
Any Sperner labeled triangulation of an $n$-simplex must contain an odd number of fully labeled elementary $n$-simplices. In particular, there is at least one.
\end{restatable}
A proof of Sperner's lemma can also be found in~\citep{Su99sperner}.

\subsection{Cake-Cutting}
\label{sec:cakecutting}

\paragraph{Model.}
A cake $\C$ refers to a divisible resource, represented by the interval $[0,1]$.
A \emph{piece} of cake refers to a finite union of disjoint subintervals of $[0,1]$.\footnote{We allow disjoint subintervals to intersect in a set of measure zero.} A \emph{connected} piece refers to a single subinterval of $[0,1]$. There is a set of agents $N = \{1, \ldots, n\}$ among whom the cake is to be divided, each of whom must be given a connected piece of the cake. A connected \emph{division} of the cake refers to a partition of the cake into $n$ disjoint connected pieces.

\paragraph{Allocation.}
We represent a connected division of the cake as a point $x=(x_1, x_2, \ldots, x_n)$ in the $(n-1)$-simplex $\Delta^{n-1}$, and will denote $x_0 = 0$. Then the leftmost piece is $[x_0,x_1]$, the next piece is $[x_1,x_1+x_2]$, and so on. The $i^\textup{th}$ piece is the subinterval $[\sum_{j < i} x_j, x_i + \sum_{j<i} x_j]$. Since $\sum_i x_i = 1$, the entire cake is divided among the agents.
We will call such a point $x \in \Delta^{n-1}$ a \emph{cut-set} of the cake. Note that any connected division can be mapped to a unique cut-set. Thus, there is a bijection between the space of connected cake divisions and $\Delta^{n-1}$.
We allocate the $i^{\textup{th}}$ piece of the cake to agent $i$.
Formally, given a cut-set $x$, we denote by $A^x = (A_i^x)_{i \in N}$ the \emph{allocation} where the $i^{\textup{th}}$ subinterval $A_i^x := [\sum_{j < i} x_j, x_i + \sum_{j < i} x_j]$ is given to agent $i$.

\paragraph{Valuations.}
Each agent $i \in N$ has a \emph{valuation function} $v_i :  \Delta^{n-1} \rightarrow \mathbb{R}$ that is continuous and maps a cut-set $x$ to a value. We require that any agent receiving an empty piece has value $0$. Thus, we enforce that $v_i(x) = 0$ whenever $x_i = 0$. The valuation profile $\mathcal V = (v_1,v_2,\dots,v_n)$ is an ordered collection of valuation functions. A connected fair division \emph{instance} is represented as $\mathcal{I} = (N, \C, \mathcal V)$.%

\paragraph{Global valuations.}
Note that an agent's valuation can depend on the entire cut-set. The three-agent instance with valuation functions $v_1(x) = x_1 \cdot (x_2 + x_3)$, $v_2(x) = - x_1 \log (1 + x_2)$, and $v_3(x) = (1 - e^{x_3}) \cdot \log(2 + x_1)$ for a cut-set $x=(x_1, x_2, x_3)$ is an example of such valuations, which we call \emph{global valuations}.

\paragraph{Local valuations.}
As a special case of global valuations, one can consider a setting where each agent $i \in N$ is associated with a \emph{local} valuation function $v_i : [0,1]^2 \rightarrow \mathbb{R}$ that is continuous and satisfies $v_i(A_i^x) = 0$ if $A_i^x$ has measure zero. Thus, for local valuations, each agent's value depends only on the piece of cake they are allocated. Local valuations are clearly a subclass of global valuations.

\paragraph{Permuted instances.}
We say two instances $\mathcal{I} = (N, \C, \mathcal V)$ and $\mathcal{I}' = (N, \C, \mathcal V')$ with local valuations $\mathcal{V} = (v_i)_{i \in N}$ and $\mathcal{V}' = (v_i')_{i \in N}$ are \emph{permutations} of each other if there is a permutation $\pi:N \rightarrow N$ so that for all $i \in N$, $v_i = v_{\pi(i)}'$, i.e., the agent valuations are permuted by $\pi$ from $\mathcal{I}$ to $\mathcal{I}'$. We can equivalently think of the agent valuations remaining unchanged and the allocation of the pieces in the cut-set being permuted.

\paragraph{Additive valuations.}
A well-studied and straightforward special case of local valuations is \emph{additive} valuation functions. Intuitively, a valuation function $v_i$ is \emph{additive} if there exists an integrable density function $f_i:[0,1] \rightarrow \mathbb{R}$ so that for any cut-set $x$, $v_i(A_i^x) = \int_{A_i^x} f_i(z) \,dz$.
Formally, a valuation function $v_i$ is \emph{additive} if it is a finite signed measure on $[0,1]$ with the Lebesgue $\sigma$-algebra. It follows that for an additive valuation function $v_i$ and two intervals $[a,b]$ and $[b,c]$, $v_i([a,c]) = v_i([a,b]) + v_i([b,c])$. Since all local valuations have zero value for measure zero sets, any such $v_i$ is absolutely continuous with respect to the Lebesgue measure. Thus the Radon-Nikodym theorem for signed measures gives the existence of a function $f_i: [0,1] \rightarrow \mathbb R$ such that $v_i(S) = \int_S f_i(z) \,dz$ where $\int \cdot \,dz$ is the Lebesgue integral. The term additive comes from the countable additivity requirement for signed measures.
Additive valuations are a subclass of local valuations with $v_i([a, b]) = \int_{a}^b f_i(z) dz$. Note that any integrable density function $f_i: [0,1] \rightarrow \mathbb R$ also gives rise to a natural additive valuation function defined as $v_i(S) = \int_S f_i(z) dz$.

\paragraph{Nonnegative valuations.}
A global/local/additive valuation function is nonnegative if it always has nonnegative value for any cut-set. For an additive valuation function, this condition is equivalent to the density function being nonnegative except for a set of measure zero.

\paragraph{Monotone valuations.}
Another commonly studied class of local valuations in cake-cutting is that of \emph{monotone} valuations, where for any subsets $S, T \subseteq \C$ such that $S \subseteq T$, we have that $v_i(S) \le v_i(T)$ for every agent $i \in N$. Observe that since the empty piece is valued at $0$, monotone valuations are a subclass of global nonnegative valuations.

\paragraph{Equitability.}
Given a connected fair division instance $\mathcal{I} = (N, \C, \mathcal V)$, a cut-set $x$ is \emph{equitable} if all the agents have equal value for the cut-set, i.e.,
\[
    v_1(x) = v_2(x) = \cdots = v_n(x) \, .
\]

For a local valuation instance, a cut-set $x$ (and the resulting allocation $A^x$) is equitable if all agents have equal value for their allocation, i.e.,
\[
    v_1(A_1^x) = v_2(A_2^x) = \cdots = v_n(A_n^x) \, .
\]

\paragraph{Happy, happier, and happiest agent(s).}
Given a connected fair division instance $\mathcal{I} = (N, \C, \mathcal V)$ with global valuations and a cut-set $x$, denote by $v_{\max}(x)$ the largest value any agent has for their piece, i.e., $v_{\max}(x) \coloneqq \max_{i \in N} v_i(x)$.

\begin{itemize}
    \item An agent $i$ is a \emph{happy agent} at $x$ if $v_i(x) = v_{\max}(x)$. Denote the set of all happy agents by $H(x) \coloneqq \{ i \in N \mid v_i(x) = v_{\max}(x) \}$.
    \item Among the happy agents, the \emph{happier agents} are the ones with the largest quantity of cake. Define $H^+(x) \coloneqq \{ k \in H(x) \mid x_k = \max_{i \in H(x)} x_i \}$ as the set of happier agents.
    \item Finally, the \emph{happiest agent} at $x$ is lexicographically the first agent in $H^+(x)$, namely, $\happiest(x) \coloneqq \arg\min_{k \in H^+(x)} k$. Given any cut-set $x$, there is a unique happiest agent $\happiest(x)$.
\end{itemize}

A cut-set $x$ is equitable if and only if all the agents are happy, i.e., $H(x) = N$. Example~\ref{eg:Happiness-Illustrated} illustrates these notions.

\begin{example}[Happy, happier, and happiest agent]
Consider an example with three agents with additive valuations with the following density functions:
\[
f_1(z) =
\begin{cases}
    5/2 & \text{if $z \in [0,2/5]$} \\
    0 & \text{otherwise}
\end{cases},
f_2(z) =
\begin{cases}
    5/4 & \text{if $z \in [1/5,3/5]$} \\
    5/6 & \text{otherwise}
\end{cases},
f_3(z) =
\begin{cases}
    5/4 & \text{if $z \in [3/5,1]$} \\
    5/6 & \text{otherwise}
\end{cases}.
\]

Consider the cut-set $x=(1/5,2/5,2/5)$ and the corresponding allocation $A^x$. Thus,  %
agent $1$ gets $[0, 1/5]$, agent $2$ gets $[1/5, 3/5]$, and agent $3$ gets $[3/5, 1]$. Then, each agent has a value $v_i(A_i) = 1/2$ for their piece of the cake, and all agents are happy. Since agent $1$ has a piece of length $1/5$, while the other two have pieces of length $2/5$ each, agents $2$ and $3$ are the happier agents. Finally, since $2$ comes lexicographically before $3$, agent $2$ is the happiest agent.
\label{eg:Happiness-Illustrated}
\end{example}

\paragraph{\myinstance{} instances.} A fair division instance $\instance$ with global valuations is a \emph{some agent nonnegative} (\myinstance{}) instance if, for every cut-set $x$, there exists an agent $i$ such that $x_i > 0$ and $v_i(x_i) \ge 0$.

It is important to note that the ordering matters for an instance to be SANN. If the ordering of the agents of a SANN instance is permuted, it may no longer remain SANN.

\begin{remark}
In a SANN instance, all agents have nonnegative value for the entire cake. This is because at the vertex $e_i$ of the $(n-1)$-simplex, agent $i$ is the only agent with $x_i > 0$, and thus has to satisfy $v_i(e_i) \ge 0$. Further, if the entire cake goes to a single agent, then that agent is the unique happiest and happier agent.
\end{remark}

Prior work~\citep{CDP13eqdivisible,AD15connected,Ch17eqdivisible} has shown that for \emph{nonnegative} additive instances and any ordering of the agents, there exists a connected equitable allocation. Example~\ref{eg:Indexing-Matters} shows that if we allow \emph{negative} additive valuations, the ordering of the agents is crucial in showing the existence of a connected equitable division.

\begin{example}[Agent ordering matters]
Consider an additive instance with three agents where
\[
f_1(z) =
\begin{cases}
    -1 & \text{if $z \in [0,1/2]$} \\
    3 & \text{if $z \in (1/2, 1]$}
\end{cases}, \quad
f_2(z) = 1,  \text{ and} \quad
f_3(z) =
\begin{cases}
    3 & \text{if $z \in [0,1/2]$} \\
    -1 & \text{if $z \in (1/2, 1]$}
\end{cases} \, .
\]

Here, $v_i([0,1]) = 1$ for all three agents. We will show the nonexistence of a connected equitable allocation for this instance for the given agent ordering.

Consider any cut-set $x$, and let $A^x$ denote the corresponding allocation. Suppose, for contradiction, that $A^x$ is equitable. If $x_1 = 0$, then $x_2$ must also be zero, else $v_2(A_2^x) > 0$. But then $x_3 = 1$, and $v_3(A_3^x) = 1$, which is not equitable. Similarly, if $x_3 = 0$, then $x_2$ must be zero, and $x_1 = 1$, which again is not equitable.

Thus, we must have $x_1 > 0$ and $x_3 > 0$. Note that agent $2$ cannot obtain negative value for any subset of the cake, hence $v_2(A_2^x) \ge 0$. If $x_1 \le 1/2$, $v_1(A_1^x) < 0$, and $A^x$ cannot be equitable. But if $x_1 > 1/2$, then $x_3 < 1/2$, and since $x_3 > 0$ by assumption, $v_3(A_3^x) < 0$, giving a contradiction. Thus, there is no connected equitable allocation in which agents are ordered $(1,2,3)$ from left to right.

The nonexistence of an equitable allocation in the above instance is due to agent $1$ disliking a prefix of the cake and agent $3$ disliking a suffix. This issue can be remedied by switching the agent ordering. That is, consider the permuted instance $(N,\C,\mathcal{V}')$ with $v_1' = v_3$, $v_2' = v_2$, and $v_3' = v_1$. Then the cut-set $x = (1/5,3/5,1/5)$ has value $3/5$ for all agents, and is hence equitable. In fact, one can verify that the permuted instance $(N,\C,\mathcal{V}')$ is a \myinstance{} instance.
\label{eg:Indexing-Matters}
\end{example}

One might wonder whether a similar impossibility, as described above, holds for the case of two agents. Prior work shows that for two agents with additive valuations, a connected equitable division always exists~\citep{CDP13eqdivisible,PW17lower}. In the concluding remarks (Section~\ref{sec:Conclusion}), we extend this existence result to agents with global valuations, as long as both agents have nonnegative value for the entire cake (but may have negative value for subsets of the cake).

In the next section, we will establish our main result: For any \myinstance{} instance, there always exists a connected equitable allocation of the cake. Note that one could also define some agent nonpositive (SANP) instances as follows, and obtain similar results.

\paragraph{SANP instances.} A fair division instance $\instance$ with global valuations is a \emph{some agent nonpositive} (SANP) instance if, for every cut-set $x$, there exists an agent $i$ such that $x_i > 0$ and $v_i(x_i) \le 0$.

For any SANP instance, one can construct a corresponding SANN instance by considering the modified valuations $v_i' = - v_i$ for all agents. Any equitable division in this SANN instance is also an equitable division in the original SANP instance, hence we focus on SANN instances in the following.

\section{Existence of a Connected Equitable Allocation}
\label{sec:existence}

We note that if some agents have positive value for the entire cake ($v_i(e_i) > 0$) while others have negative value for the entire cake ($v_i(e_i) < 0$), there may not exist an equitable allocation --- connected or otherwise. Consider an additive instance with two agents, with density functions $f_1(z) = -1$ and $f_2(z)=1$. It is easy to see that if both agents get nontrivial pieces, then agent 1 gets a negative value while agent 2 gets a positive value. If one of the agents gets a trivial allocation, then it gets a zero value while the other gets a nonzero value.

Therefore, we restrict attention to instances where all agents have nonnegative value for the entire cake, i.e., $v_i(e_i) \ge 0$.\footnote{If an agent has value 0 for the entire cake, then assigning the entire cake to this agent is a connected equitable allocation.} Recall that this condition is satisfied by all \myinstance{} instances. The case where all agents have nonpositive value is symmetric.

In this section, we will show that any \myinstance{} instance admits a connected equitable allocation. An easy example of a \myinstance{} instance is one where all valuations are additive and nonnegative. In Section~\ref{sec:nonmonotone}, we will discuss examples of instances that may be negative-valued, yet still satisfy the \myinstance{} property and therefore allow for a connected equitable allocation.

To show the existence of a connected equitable allocation, we will use Sperner's lemma (Proposition~\ref{lem:Sperners}). We will begin by establishing that a triangulation that is labeled according to the index of the happiest agent satisfies the conditions of Sperner labeling. In our proof, we only need the set of happier agents, as the happiest agent only comes into play for breaking ties.

\begin{restatable}{lemma}{labeling}
\label{lem:labeling}
Given any triangulation $T$ of the $(n-1)$ simplex $\Delta^{n-1}$ and a \myinstance{} instance $(N, \mathcal C, \mathcal V)$, suppose each vertex $x$ of the triangulation is labeled with the index $\happiest(x)$ of the happiest agent under the corresponding cut-set $x$. Then, the resulting labeling is a Sperner labeling.
\end{restatable}
\begin{proof}
Recall that $e_i \in \R^n$ has $1$ in coordinate $i$ and $0$ everywhere else. We need to show two properties:
\begin{enumerate}
    \item Each vertex of the simplex $\Delta^{n-1}$ has a different label.
    \item The facet opposite the vertex labelled $i$ of $\Delta^{n-1}$ has no vertex of the triangulation labeled $i$.
\end{enumerate}

To see why Property 1 holds, recall that at a vertex $x = e_i$, by definition for \myinstance{} instances, agent $i$ has a nonnegative value, while every other agent has a trivial allocation and hence zero value. Thus, the only happier agent is $i$. Each vertex is thus labeled with a unique index.

Let us now show that Property 2 holds. Since $x_i=0$ for any cut-set $x$ in the facet opposite vertex $e_i$, by definition of \myinstance{} instances, there exists an agent $j \neq i$ with $x_j > 0$ and $v_j(x) \ge 0$. Since $x_i = 0$ and $v_i(x) = 0$, agent $i$ cannot be happier. Thus, any vertex on the facet opposite the vertex $e_i$ does not contain the label $i$.
\end{proof}

Sperner's lemma implies that there exists a fully labeled elementary simplex.
The vertices of this elementary simplex correspond to $n$ distinct cut-sets such that a different agent is happiest at each cut-set. Using finer and finer triangulations, we demonstrate that these cut-sets can be arbitrarily close to each other. This gives us a sequence of increasingly smaller fully labeled elementary simplices. In each of these simplices, each agent is happiest at exactly one vertex. Because the space of connected divisions is compact, there is a single vertex (and a corresponding cut-set) where all agents are happy, which implies that all agents have the same value, and the cut-set is hence equitable.
We formalize the notion of ``finer and finer'' triangulations in the following definition.

\paragraph*{Vanishing triangulations.}

A sequence of triangulations $\{T^k\}_{k \in \mathbb N}$ of $\Delta^{n-1}$ is said to be \emph{vanishing} if for all $\eps > 0$, there exists $K \in \mathbb N$ such that for all $k' \ge K$ and for all $\delta \in T^{k'}$,  we have $\diameter(\delta) < \eps$. Here, $\diameter(\delta) = \max_{x, y \in \delta} \|x - y\|_2$.

Such a sequence can be obtained, for example, using barycentric subdivision~\cite{Su99sperner, SH18burntcake}. We then use the following lemma to argue that, if we have a fully labeled elementary simplex for each triangulation, then we can obtain a single cut-set that is equitable.

\begin{restatable}{lemma}{cutsetsequence}
\label{lem:cutsetsequence}
Let $\{T^k\}_{k \in \mathbb N}$ be a sequence of vanishing triangulations of $\Delta^{n-1}$. Let $\{\delta^k\}_{k \in \mathbb N}$ be a sequence of simplices such that $\delta^k \in T^k$ for all $k \in \mathbb N$. There exist a subsequence $\{\delta^t\}_{t \in \mathcal T}$ and an $x^* \in \Delta^{n-1}$ such that for all sequences $\{x^t\}_{t \in \mathcal T}$ with $x^t \in \delta^t$, we have $x^t \to x^*$.
\end{restatable}

\begin{proof}
Let $\{y^k\}_{k \in \mathbb N}$ be any sequence of points such that $y^k \in \delta^k$ for all $k \in \mathbb N$. By the Bolzano-Weierstrass theorem, there exists a subsequence $\{y^t\}_{t \in \mathcal T}$ such that $y^t \to x^*$ for some point $x^*$. We show that the subsequence of simplices $\{\delta^t\}_{t \in \mathcal T}$ and the point $x^*$ satisfy the required properties. Fix any sequence $\{x^t\}_{t \in \mathcal T}$ with $x^t \in \delta^t$, and any $\eps > 0$. Since $y^t \to x^*$, there exists a $\tau$ such that for all $t \ge \tau$, $\|y^t - x^*\|_2 \le \eps/2$. Since $\{T^k\}$ is a vanishing sequence of triangulations, there exists a $K \in \mathbb N$ such that for all $t \ge K$ and for all $x^t, y^t \in \delta^t$ with $\delta^t \in T^t$, it holds that $\|x^t - y^t\|_2 \le \eps/2$.

Thus, for all $t \ge \tau+K$, it holds that
\[
\|x^t - x^*\|_2 \le \|x^t - y^t\|_2 + \|y^t - x^*\|_2 \le \eps
\]
which proves the claim.
\end{proof}

\begin{restatable}{theorem}{existence}
\label{thm:existence}
Any \myinstance{} or SANP instance $(N, \C, \mathcal V)$ admits a connected equitable allocation.
\end{restatable}
\begin{proof}
For SANP instances, we consider the corresponding SANN instance obtained with valuations $v_i' = - v_i$ in what follows.
Consider a sequence of vanishing triangulations $\{ T^k \}_{k \in \mathbb N}$ of the $(n-1)$-simplex $\Delta^{n-1}$. Let $\delta^k$ be the fully labeled elementary simplex of triangulation $T^k$ obtained from Sperner's lemma.

Applying \cref{lem:cutsetsequence} to the sequence of simplices $\{\delta^k\}_{k \in \mathbb N}$, we obtain a subsequence of simplices $\{\delta^t\}_{t \in \mathcal T}$ and a point $x^* \in \Delta^{n-1}$.
Fix any agent $i \in N$. Since $\delta^t$ is fully labeled, agent $i$ is the happiest agent of exactly one vertex of $\delta^t$, say $x^{i,t}$.
In particular, $i$ is a happy agent at this vertex $x^{i,t}$, and thus $v_i(x^{i,t}) \ge v_j(x^{i,t})$ for all other agents $j$.
Since the sequence $\{x^{i,t}\}$ satisfies $x^{i,t} \in \delta^t$, we have (by \cref{lem:cutsetsequence}) that $x^{i,t} \to x^*$. Since the valuation functions are continuous, we also have $v_i(x^*) \ge v_j(x^*)$ for all other agents $j$ at this limit point $x^*$, and thus $i$ is happy at $x^*$.

Since the choice of agent $i$ was arbitrary, and since the point $x^*$ is independent of the choice of agent $i$,
the above argument shows that every agent is happy at $x^*$. Thus $x^*$ must be equitable.
\end{proof}

The proof of Theorem~\ref{thm:existence} used the property that an agent who is happy at each cut-set in the sequence continues to be happy in the limit cut-set. It is relevant to note that this property may not hold for \emph{happier} agents. In other words, an agent that is happier in a sequence of cut-sets may not be happier at the limit cut-set. This is because the concept of happier agent is defined only with respect to the set of happy agents. %
It is possible for agent $i$ with a small piece of the cake to be happier in a sequence of cut-sets by being the only happy agent in each of those cut-sets, but at the limit point, there could be multiple happy agents with larger pieces of cake, making $i$ no longer a happier agent. In essence, an agent who is not happy in a sequence of cut-sets may become happy in the limit cut-set, which could change the set of happier agents.

We obtain the existence of connected equitable divisions for the following valuation class immediately, since they are easily seen to be \myinstance{} instances.

\begin{corollary}
    Every instance where all agents have global valuation functions that are nonnegative admits a connected equitable division.
    \label{cor:indexing}
\end{corollary}

In particular, since the above result holds for \emph{every} such instance, it also holds for instances where the valuations are permuted. Thus, in a nonnegative instance, for every ordering of the agents, there exists a connected equitable division which allocates the cake in that order.
Since nonnegative additive valuations are a subclass of nonnegative valuations, we recover the following result from prior work.

\begin{corollary}
Every permutation of an instance with nonnegative additive valuation functions admits a connected equitable division.
\label{cor:additive}
\end{corollary}

As an aside, SANN does not completely characterize the class of instances to which this proof technique applies, since the weaker condition of admitting a Sperner labeling suffices for applying Sperner's Lemma.

\section{Subclasses of Additive Valuations}
\label{sec:nonmonotone}

In this section, we will identify several subclasses of additive valuations to which~\Cref{thm:existence} can be applied. For an additive valuation $v_i$ with density function $f_i$, we define the \emph{cumulative distribution function} (or cdf) $F_i(t) = \int_{z=0}^t f_i(z) \, dz$. We will show that the subclasses considered here satisfy the \myinstance{} condition.
We assume in what follows that $v_i(e_i) \ge 0$ for all $i$, and the density (and thus the valuation) is allowed to be negative in $[0,1]$.

\paragraph*{Value-ordered instances.} A fair division instance with additive valuations $(N, \C, \mathcal V)$ is \emph{value-ordered} if
 for any $t \in [0,1]$ and $i \in [n-1]$, $F_i(t) \ge F_{i+1}(t)$ for $i \in [n-1]$. That is, the agents are ordered so that any initial piece of the cake has a greater value for an agent on the left.

We will now show that any value-ordered instance satisfies the \myinstance{} property.

\begin{restatable}{lemma}{value-ordered}
\label{lem:value-ordered}
A value-ordered instance is a \myinstance{} instance when all agents have non-negative value for the entire cake, i.e., $v_i(e_i) \ge 0$ for all agents $i$.
\end{restatable}

\begin{proof}
Consider any allocation $A^x$ from the cut-set $x = (x_1, \ldots,x_n)$. %
We will show that for each $i \in N$, the total value obtained by agents $\{1, \ldots, i\}$ is at least $F_i(x_1+x_2+\dots+x_i)$. Note that by invoking the claim for $i = n$, we get that the total value obtained by all the agents is at least $F_n(1)$, which is nonnegative by assumption. Hence, either every agent receives zero value for their piece, or some agent must receive positive value, completing the proof of the lemma.

We will prove the above claim by induction. For the base case, we have $i=1$. Then, the value obtained by agent 1 is $\int_{0}^{x_1} f_1(t) dt$, which is exactly $F_1(x_1)$. Assume the claim is true for some $i$. Then the total value obtained by agents $j \le i+1$ is:
\begin{align*}
\sum_{j \le i} & \biggl(F_j \Bigl( \sum_{k \leq j} x_k \Bigr) - F_j \Bigl(\sum_{k \leq j-1} x_k \Bigr)\biggr) + \biggl(F_{i+1} \Bigl( \sum_{k \leq i+1} x_k \Bigr) - F_{i+1} \Bigl( \sum_{k \leq i} x_k \Bigr) \biggr) \\
& \ge F_i \Bigl( \sum_{k \leq i} x_k \Bigr) + \biggl(F_{i+1} \Bigl( \sum_{k \leq i+1} x_k \Bigr) - F_{i+1} \Bigl( \sum_{k \leq i} x_k \Bigr) \biggr)\\
& \ge F_i \Bigl( \sum_{k \leq i} x_k \Bigr) + F_{i+1} \Bigl( \sum_{k \leq i+1} x_k \Bigr) - F_{i} \Bigl( \sum_{k \leq i} x_k \Bigr)\\
&  = F_{i+1} \Bigl( \sum_{k \leq i+1} x_k \Bigr),
\end{align*}
where the first inequality is by the induction hypothesis, and the second inequality is by the property of value-ordered instances. This completes the proof.
\end{proof}

An additive instance is said to have \emph{identical} valuations if all agents have the same density function. That is, for every pair of agents $i \neq j$, we have $f_i = f_j$. It is easy to see that identical valuations are value-ordered and thus satisfy the \myinstance{} property.

\begin{restatable}{corollary}{identical}
\label{cor:identical}
An additive instance with identical valuations is a \myinstance{} instance.
\end{restatable}

\paragraph*{Split-cake instances.} We say an agent $i$ has a \emph{split-cake} valuation if there exists an interval $I_i = [l_i, r_i] \subseteq [0,1]$ where the agent has nonnegative density, and has nonpositive density outside the interval. That is, $f_i(z) \le 0$ for $z \in [0,l_i) \cup (r_i,1]$ and $f_i(z) \ge 0$ for $z \in [l_i, r_i]$. An instance with split-cake agents is a split-cake instance. We will show that every split-cake instance admits an ordering of the agents that is a \myinstance{} instance, which then admits a connected equitable division.

\begin{restatable}{lemma}{split-cake}
\label{lem:split-cake}
A split-cake instance admits a \myinstance{} permutation when all agents have non-negative value for the entire cake, i.e., $v_i(e_i) \ge 0$ for all agents $i$.
\end{restatable}

\begin{proof}
The proof follows from three steps. First, we show that for every agent $i$ with a split-cake valuation, there exists a threshold $\theta_i$ such that the cumulative value $F_i(z) \le 0$ for $z \le \theta_i$, and $F_i(z) \ge 0$ for $z \ge \theta_i$. Then, we order the agents by thresholds so that $\theta_1 \le \theta_2 \le \ldots \le \theta_n$. Lastly, we show that for any allocation of the cake, there exists an agent $i$ with a nontrivial allocation whose allocation contains the point $\theta_i$. The proof follows, since if agent $i$ receives the piece $[a,b]$ with $b > a$ where $\theta_i \in [a,b]$, then $F_i(b) \ge 0 \ge F_i(a)$, and hence agent $i$ gets a nontrivial piece with nonnegative value $F_i(b) - F_i(a)$.

Fix an agent $i$. To show the existence of a threshold $\theta_i$, note that $F_i(z)$ is nonincreasing in the interval $[0,l_i]$, nondecreasing in $[l_i,r_i]$, and nonincreasing thereafter. Thus, $F_i(l_i) \le 0$, and since we assume $F_i(1) \ge 0$ (i.e., every agent has nonnegative value for the entire cake), $F_i(r_i) \ge 0$. We set $\theta_i$ to be any point in $[l_i,r_i]$ for which $F_i(\theta_i) = 0$. Since $F_i(z)$ is nondecreasing in $[l_i,r_i]$, we obtain $F_i(z) \le 0$ for $z \le \theta_i$, and $F_i(z) \ge 0$ for $z \ge \theta_i$, as required.
Note that if $f_i(0) > 0$, then $\theta_i = 0$.

We then order the agents so that $\theta_1 \le \theta_2 \le \ldots \le \theta_n$. Note that $0 \le \theta_1 \le \theta_n \le 1$. For any cut-set $x = (x_1, \ldots,x_n)$, let $S = \{i_1, \ldots, i_k\}$ be the set of agents that receive nontrivial pieces, and assume these are ordered by the threshold, i.e., $\theta_{i_j} \le \theta_{i_{j+1}}$ for $j \in [k-1]$. Let $i_{j^*}$ be the first agent so that $\theta_{i_{j^*}} \le \sum_{j \le j^*} x_{i_j}$. Such an agent always exists, since $\theta_{i_{j'}} > \sum_{j \le j'} x_{i_j}$ cannot hold for all agents, as the right hand term is $1$ for agent $i_k$. Then $\theta_{i_{j^*}}$ falls in the piece allocated to agent $i_{j^*}$, and agent $i_{j^*}$ gets a nontrivial piece with nonnegative value, as required.
\end{proof}

A special case of split-cake instances is \emph{single-peaked} instances. An additive valuation $v_i$ is \emph{single-peaked} if  there exists a peak $p_i \in [0,1]$ so that, for any $t \le s \le p_i$, $f_i(t) \le f_i(s)$, and for any $p_i \le s \le t$, $f_i(s) \ge f_i(t)$. Thus there is a single peak $p_i$ for the density function such that the function is increasing before $p_i$ and decreasing after $p_i$. An instance is a single-peaked instance if all agent valuations are single-peaked. Single-peaked instances have been considered in cake-cutting literature before~\cite{wang2019cake, bhardwaj2020fairness}. However, these definitions require the density functions to be linear away from the peak, and are hence a restricted case of our model.

\begin{restatable}{corollary}{single-peaked}
\label{cor:single-peaked}
A single-peaked instance admits a \myinstance{} permutation when all agents have non-negative value for the entire cake, i.e., $v_i(e_i) \ge 0$ for all agents $i$.
\end{restatable}

\paragraph*{An instance with no \myinstance{} permutation.} An obvious question is if possibly all additive instances---even allowing for negative values---have at least one \myinstance{} permutation. This
would then imply that for these instances, there always exists a connected equitable allocation for some permutation of the agents. Below, we give an example to show that this is not the case.

\begin{example}[Instance with no \myinstance{} permutation]
Consider a three-agent instance with cdfs as shown in~\Cref{fig:non-example}.

\begin{figure}[!h]
\centering
\begin{tikzpicture}
  \begin{axis}[
    xlabel={$x$},
    ylabel={$F(x)$},
    axis lines=middle,
    xmin=0, xmax=1.1,
    ymin=-1, ymax=2.5,
    xtick={0,0.1,...,1},
    ytick={-1,0,1,2},
    width=15cm,
    height=5cm,
    clip=false, %
    axis on top, %
    legend style={at={(0.5,-0.15)},anchor=north}, %
    legend cell align=left, %
    legend columns=3, %
      /tikz/every even column/.append style={column sep=0.5cm} %
  ]

  \addplot[mark=none,solid,purple] coordinates {
    (0,0) (0.1,-1) (0.2,2) (0.8,-1) (0.9,2) (1,1)
  };
  \addlegendentry{$F_1(x)$}

  \addplot[mark=none,dashed,black] coordinates {
    (0,0) (0.2,-1) (0.3,2) (0.7,-1) (0.8,2) (1,1)
  };
  \addlegendentry{$F_2(x)$}

  \addplot[mark=none,dotted,teal,line width=1.1pt] coordinates {
    (0,0) (0.4,-1) (0.7,2) (1,1)
  };
  \addlegendentry{$F_3(x)$}

  \end{axis}
\end{tikzpicture}
\caption{A three-agent instance with no \myinstance{} permutation. For every permutation of the agents, there is an allocation where every agent with a nontrivial piece has negative value.}
\label{fig:non-example}
\end{figure}
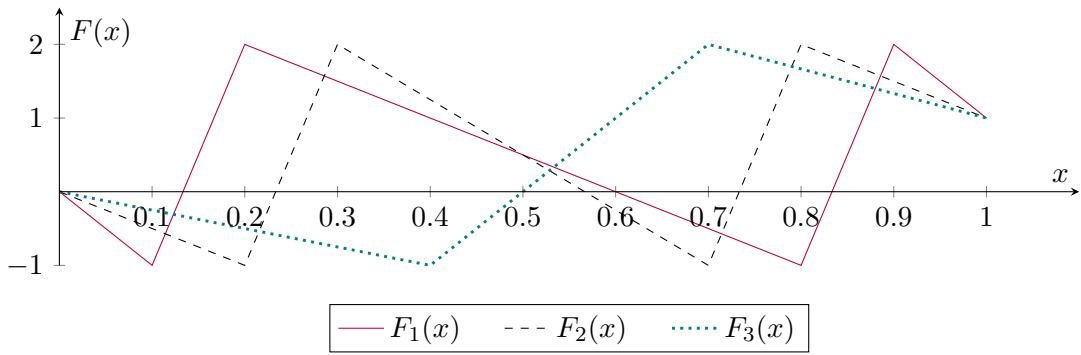

Thus the density functions in the instance are:

\[
f_1(x) =
\begin{cases}
    -10 & \text{if $x \in [0,0.1]$} \\
    +30 & \text{if $x \in (0.1,0.2]$} \\
    -5 & \text{if $x \in (0.2,0.8]$} \\
    +30 & \text{if $x \in (0.8,0.9]$} \\
    -10 & \text{if $x \in (0.9,1]$}
\end{cases},
\qquad
f_2(x) =
\begin{cases}
    -5 & \text{if $x \in [0,0.2]$} \\
    +30 & \text{if $x \in (0.2,0.3]$} \\
    -7.5 & \text{if $x \in (0.3,0.7]$} \\
    +30 & \text{if $x \in (0.7,0.8]$} \\
    -5 & \text{if $x \in (0.8,1]$}
\end{cases},
\]

\[
f_3(x) =
\begin{cases}
    -2.5 & \text{if $x \in [0,0.4]$} \\
    +10 & \text{if $x \in (0.4,0.7]$} \\
    -10/3 & \text{if $x \in (0.7,1]$}
\end{cases}.
\]

The following table shows that for every ordering of the agents, there exists an allocation where every agent with a nontrivial piece has negative value.

\begin{table}[!h]
\centering
    \begin{tabular}{|l|l|l|}
    \hline
    \centering Agent order & {\centering Allocation} & {\centering Agent values} \\ \hline
    agent 1, agent 2, agent 3 & $(0.8,0.2,0)$ & $(-1,-1,0)$ \\ \hline
    agent 1, agent 3, agent 2 & $(0.8,0,0.2)$ & $(-1,0,-1)$ \\ \hline
    agent 2, agent 1, agent 3 & $(0.7,0,0.3)$ & $(-1,0,-1)$ \\ \hline
    agent 2, agent 3, agent 1 & $(0.7,0.3,0)$ & $(-1,-1,0)$ \\ \hline
    agent 3, agent 1, agent 2 & $(0,0.8,0.2)$ & $(0,-1,-1)$ \\ \hline
    agent 3, agent 2, agent 1 & $(0.3,0.7,0)$ & $(-0.75,-1,0)$ \\ \hline
    \end{tabular}
\end{table}

The issue here is that, for any ordering of the agents, the coloring of the simplex as described in~\Cref{sec:existence} using the notions of happy, happier, and happiest agents does not give us a Sperner coloring, since in the face opposite a vertex $e_i$, the agent $i$ could be the happiest agent, despite obtaining a trivial allocation. Hence, we do not obtain the existence of a fully-labeled elementary simplex.

For this particular instance, there is, however, an equitable allocation. Consider the cut-set $x = (48/230, 45/230,137/230)$. It can be verified that for this allocation, each agent gets a value of $45/23$ (which is almost $2$, the maximum possible).
\label{eg:non-example}
\end{example}

\section{Conclusion}
\label{sec:Conclusion}

We presented a simple proof showing the existence of a connected equitable cake division. With nonnegative additive valuations, our proof shows that for any ordering of agents, a connected equitable allocation exists. However, when the density functions are allowed to be negative, re-ordering the agents may preclude a connected equitable division, as demonstrated in Example~\ref{eg:Indexing-Matters}. This example involved three agents, prompting the question of whether a similar limitation applies to the case of two agents. In Proposition~\ref{prop:twoagentsnonmonotone}, we prove that if both agents have global valuations with nonnegative value for the entire cake, then for any ordering of the agents, there exists a connected equitable division. We note that similar proofs for the two-agent case with additive valuations have been given earlier as well~\citep{CDP13eqdivisible,PW17lower}; our proof extends this to agents with global valuations, as long as both agents have nonnegative value if they are given the entire cake.

\begin{restatable}{prop}{twoagentsnonmonotone}
\label{prop:twoagentsnonmonotone}
Any fair division instance with two agents with global valuations such that $v_i(e_i) \ge 0$ for $i \in \{1,2\}$ admits a connected equitable allocation.
\end{restatable}
\begin{proof}
Let $t \in [0,1]$ denote the position of the single cut. The valuation functions $v_1$ and $v_2$ can, then, be equivalently represented as continuous functions of the single parameter $t$. Consider the function $h(t) = v_1(t) - v_2(t)$. For $t = 0$, this function is nonpositive (since $v_1(\emptyset) = 0 $ and $v_2([0,1]) \geq 0$) and for $t = 1$, this function is nonnegative (since $v_1([0,1]) \ge 0$ and $v_2(\emptyset) = 0$). It follows by the intermediate value theorem that for some $t \in [0,1]$, $v_1(t) = v_2(t)$, and this is a connected equitable division.
\end{proof}

A natural question for future work is to explore the limits of \myinstance{} instances. We believe there may be other natural classes of instances that satisfy this condition. We have seen there are additive instances with three agents with no \myinstance{} permutation. However, do there still exist connected equitable divisions for some permutation of the agents, possibly for all additive valuations? One could ask the same question about global valuations. We do not know of any instance where all agents have nonnegative value for the entire cake (and possibly negative value for subsets), and a connected equitable division does not exist for any permutation of the agents.

The difficulty in extending our framework to instances where agents may have negative values for subsets, and the importance of \myinstance{} permutations, is illustrated by the following. As before, we could label each vertex by the index of the highest-valued agent (breaking ties lexicographically). However, this does not give a Sperner labeling, since at a face, an agent with an empty piece could have the highest value (this is precisely what the \myinstance{} property avoids). Another approach could be to label each vertex by the index of the highest-valued agent from those that are assigned nonempty pieces (again, breaking ties lexicographically). This is a Sperner labeling, so a fully-labeled simplex always exists. However, the limit of a sequence of non-empty pieces corresponding to a fixed index $i$ may be an empty piece, and hence, a fully labeled point may not exist.

Finally, while our work focuses on existence results, computational questions are also meaningful. Can one obtain a connected, approximately equitable allocation efficiently, for general additive valuations? Prior work showed that there is no finite algorithm for finding an exact connected equitable division that is also proportional, even for nonnegative additive valuations~\citep{CP12computability}. Their work also gave an algorithm with query complexity $\mathcal O \!\left(n \log (n/\varepsilon)\right)$ for computing a proportional $\varepsilon$-approximate connected equitable division.
Since a permutation of a SANN instance could possibly not be a SANN instance, our work introduces the additional computational question of finding a valid permutation of the agents that satisfies SANN --- all previous results on the existence of a connected equitable division held irrespective of the permutation of the agents. Thus, beyond finding a single SANN permutation  (or one that admits an equitable allocation), an intriguing question is that of counting the number of permutations of a given instance that is SANN (or admit equitable allocations).

\section*{Acknowledgments}
UB acknowledges support from the Department of Atomic Energy, Government
of India, under project no. RTI4001. RV acknowledges support from
DST INSPIRE grant no. DST/INSPIRE/04/2020/000107 and SERB grant no. CRG/2022/002621.

\bibliographystyle{gamma}
\def\bibfont{\small}
\bibliography{Ref}
\end{document}